# Towards a Multi-Agent System Architecture for Supply Chain Management

Carlos R. Jaimez-González and Wulfrano A. Luna-Ramírez

Departamento de Tecnologías de la Información, Universidad Autónoma Metropolitana - Cuajimalpa, Av. Constituyentes No. 1054, Col. Lomas Altas, C.P. 11950, México D.F.
`{cjaimez, wluna}@correo.cua.uam.mx`

**Abstract.** Individual business processes have been changing since the Internet was created, and they are now oriented towards a more distributed and collaborative business model, in an e-commerce environment that adapts itself to the competitive and changing market conditions. This paper presents a multi-agent system architecture for supply chain management, which explores different strategies and offers solutions in a distributed e-commerce environment. The system is designed to support different types of interfaces, which allow interoperating with other business models already developed. In order to show how the entire multi-agent system is being developed, the implementation of a collaborative agent is presented and explained.

## 1 Introduction

The great amount of information and the new technologies increase the client expectations about services and their costs, as well as the global competition makes enterprise leaders look for new ways of handling businesses. Companies are not in a position to trust in static business strategies, but they have to be capable of facing environments that are uncertain and change rapidly; such as changes in the bank currency values, political situations, delays in the delivery of production materials, broken relationships with suppliers, failure in production facilities, workers absence, cancelations or changes of client orders, etc.

Companies have to carry out a series of activities, such as obtaining materials, manufacturing products, storage of products, sales and delivery of products, services to the client, among others. All these activities have to be carried out as if they were a dynamic process, in such a way that there is a balance between them. In a company, this is precisely the main task of the Supply Chain Management (SCM); which is in charge of negotiating with suppliers to obtain all the necessary materials for production, handling client orders, controlling the inventory, establishing the amount of time dedicated for manufacturing and delivering finished products, etc.

Taking into consideration the market globalization, it is common that companies have distributed businesses, where suppliers and clients are located around the world. The development of the information technologies have produced that organizations use the Internet to participate in e-commerce activities; this way they reduce administrative and transactional costs, increase their profit, and interact with a greater number of business partners in different geographical locations. The Internet has produced the change of individual business processes, towards a more distributed and collaborative model. In order to be capable to handle this model, companies need a solution that allows them participate in e-commerce environments. This solution has to include a system that helps in the decision-making, and at the same time adapts to the changes; such system can collect and process information from a great number of heterogeneous sources, as well as help making decisions more precisely in competitive and changing market conditions. Due to the need of designing strategies for coordinating and integrating business entities within these environments, one of the objectives of this work is the development of strategies with agent technology for e-commerce. For the implementation and testing, it is needed to develop a system for supply chain management in the environments mentioned.

This paper presents a multi-agent system architecture for supply chain management, which is based on multiple collaborative agents, explores different strategies of the production process globally, and offers solutions for managing supply chains in distributed e-commerce environments. The system is designed to support different types of interfaces, which allows interacting with the existing business models of others participants, through the management or integration of the cooperative supply chain. An implementation of a collaborative agent is presented and explained, so that it can be observed how the entire system is being developed. Because the electronic market is a recently established business model and it is conceived as an activity for solving cooperative distributed problems, the design of a system for supply chain management has become more important than ever. It should be noticed that the work presented in this paper, is an extension of an initial proposal for developing a multi-agent system with collaborative agents, which is described in [1].

The rest of the paper is organized as follows. Section 2 provides preliminaries and state of the art for SCM, its progress, the multi-agent approach, and an experimentation platform. In section 3 it is presented the architecture of the system and its six collaborative agents. Some sales and production strategies for e-commerce agents are presented in section 4. Finally, we provide conclusions and future work in section 5.

## 2      Preliminaries and State of the Art

This section provides some preliminaries, and summarizes the state of the art in the field of the SCM, and the decision support systems for SCM. In particular, the tendency in these areas is to move from the static business processes, towards dynamic and distributed models. The decision support systems for SCM are currently designed as multi-agent systems to support such models. Many research groups around the world dedicate their work to explore several problems in the domain of the SCM, and have carried out studies and experiments on the simulation platform called *Trading Agent Competition – Supply Chain Management* (TAC SCM), which is used widely in research, and it is described at the end of this section.

### 2.1 SCM Progress

SCM is a complex process, which includes a variety of interrelated activities, such as negotiation with suppliers to obtain materials, competition for client orders, inventory management, production schedule, delivery of products to clients, etc. SCM concepts have been used by companies since the beginning of the 20$^{th}$ century, even in the literature we can find these concepts since 1950 [2]. In the 80's the idea of automate business processes through SCM was very popular, however the experts involved, isolated every entity of the supply chain as a static process, separated from the rest. Some of the work carried out towards the end of the 90's concentrated in solving specific separated areas of the SCM [3], [4]. In [5], it started to conceive the SCM problem as a dynamic environment and as an integrated process with constraints [6].

With the creation of the WWW, e-commerce systems have become extremely popular, mainly in the last decade. There are e-commerce integrated models, which include suppliers, clients, commercial partners, e-commerce agents, etc., within a global electronic environment [7]. Recently, there have been proposals for system architectures to support the participation in e-commerce [8]. Many researchers agree that the architecture of a decision support system for SCM must be agile in order to compete in the dynamics of electronic markets, as well as easily configurable, adaptable to several businesses, and support several protocols of different commercial environments. The multi-agent approach has demonstrated to be the most appropriate to fulfill these requirements [9], [10], [11], [12].

### 2.2 Multi-agent approach

Agent technology has become the most popular tool to design distributed systems for supply chain management, because it provides a dynamic and adaptable way of managing separately every piece of the chain. Agent-based supply chain management systems can respond rapidly to internal or external changes through decision-making mechanisms. A detailed review of some agent-based systems, which were designed for industrial purposes, can be found in [13].

Another advantage of designing the SCM solution as a multi-agent system is that allows separating the different tasks of the SCM, exploring them independently, and analyzing them as a whole. This feature is particularly important because it allows focusing on two important pieces of the supply chain: the demand, which refers to the sales to clients; and the production, which involves getting the raw materials and manufacturing the products. In the case of the sellers, the main problem they face when dealing with supply chains is to decide what offers they will place to their clients, what prices they will give them, when to sell their products, etc. The aim is to increase their profit; but the task is not easy to solve in the e-commerce context, where prices are established dynamically.

It is important to note that there have been proposals using different approaches to design agent-based decision-making systems for SCM [11], and also there have been developments of agent-oriented architectures for e-commerce [7]. There has been also an integration of a multi-agent system for SCM in [14]. A description of recent projects that use agents in the context of SCM is provided in [15].

### 2.3 The Experimentation Platform

Real companies do not implement the proposed solutions immediately; instead they wait until solutions have been fully tested through a series of experiments with the production and sales strategies proposed. There have been many attempts to create simulation and experimentation tools and platforms for SCM, which allow studying and testing the different algorithms and strategies. Among these attempts, there is an experimentation platform called the *Trading Agent Competition Supply Chain Management* (TAC SCM) [16], which allows testing SCM systems, because it encapsulates many of the problems that can be faced in real SCM environments, such as time constraints, unpredictable opponents, limited production capacity, late deliveries, missing components or materials, etc. This platform will be used for the experimentation of the system we proposed; specifically for testing the performance solution of the complete system, as well as for the production and sales strategies implemented.

This platform was designed by the Carnegie Mellon University (CMU) and the Swedish Institute of Computer Science (SICS) in 2003, as part of the International Trading Agent Competition [17]. The platform simulates a game, in which trading agents compete in the context of SCM, such that they can evaluate their algorithms and proposals. The agents that participate in this competition are developed by different research groups around the world.

During the game there are six agents that compete among them in a simulation that lasts for 220 days, with a real duration of 55 minutes. Each agent is a manufacturer of computers, which assembles them from their basic components, such as CPUs, motherboards, RAM memories, and hard disks. The CPUs and motherboards are available in two families of different products: IMD and Pintel. A Pintel CPU only works with a Pintel motherboard, while an IMD CPU can only be incorporated to an IMD motherboard. The CPUs are available in two speed models: 2.0 and 5.0 GHz; the memories are produced in two sizes: 1 and 2 GB; and the hard disks are available in two sizes: 300 and 500 GB. There are ten different components in total, which can be combined in order to have a total of 16 different models of computers. All the agents in the game need to obtain these components from 8 different suppliers; and at the same time the agents need to secure client orders every day, assemble a number of computers of the required model, and subsequently deliver the finished products. All the agents start the game with no money in their bank accounts, with no components of finished components in their inventory, and with no pending client orders. Every computer model requires a different number of cycles to be assembled, and an agent has a limited production capacity of 2000 cycles a day. The TAC server simulates the suppliers, clients and bank; and also provides production and storage services to all the six agents. The agent who accomplishes the greatest profit at the end of the game is the winner. The rules of the game, as well as the required software to carry out experimentation with this platform can be downloaded directly from [17]. Figure 1 illustrates all the tasks that an agent has to carry out every day during a TAC SCM game, such as dealing with suppliers, customers, managing the inventory, programming the production and delivery schedules, etc.

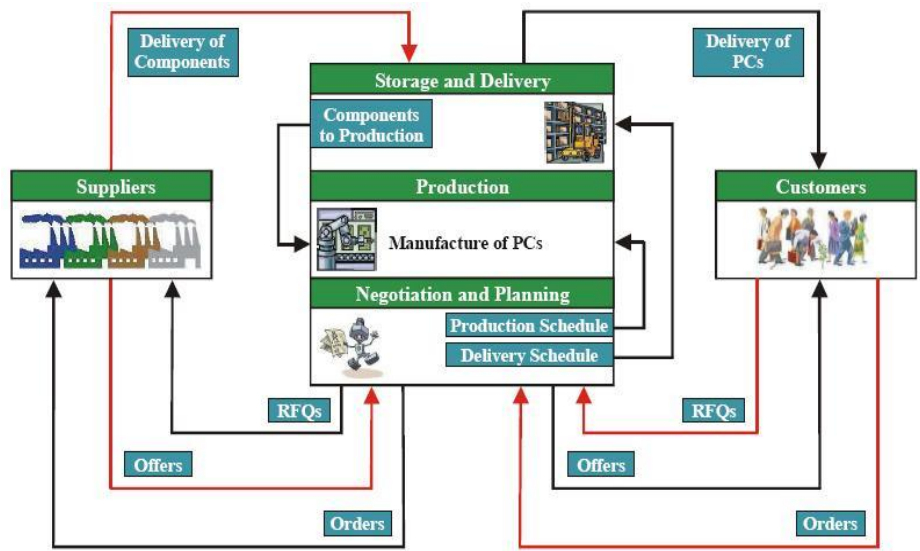

**Fig. 1.** Daily tasks of an agent in TAC SCM.

## 3   System Architecture

This section describes the system architecture, and the six agents that are part of it. The system is being developed using the principles of the multi-agent systems. Using this approach allows us studying several of the problems that can be faced in the SCM domain. Although we planned that the system competes in the game organized by the TAC SCM, its architecture will be generic and configurable, in such a way that can be adapted to similar environments, as well as modifiable to meet the particular requirements of a specific business. Tackling this problem with a multi-agent approach allows us exploring different problems of the SCM separately. We also propose to use a web service based architecture through which we can interoperate with other systems or agents developed with other technologies or programming languages.

In a commercial environment, the system should produce certain decisions in the acquisition of materials, sales, production, and delivery of products. In order to make these decisions, it is needed to complete several tasks, which can be linked to the result of others or be independent. Because there are time constraints for decision-making, the independent tasks must be carried out in parallel to save time. Trying to solve this problem with a multi-agent approach is adequate to meet these requirements. Through the coordination and collaboration, the agents are capable of managing the distributed activities in the SCM. The multi-agent approach is a "*natural way of modularizing complex systems*" [15], such as the systems for SCM.

The system is composed of six agents; five of them correspond to each of the entities of the supply chain: Sales Manager, Supply Manager, Inventory Manager, Production Manager, Delivery Manager, and the sixth agent is the Coordinator, who is in charge of coordinating the performance of the whole system, as well as communicating with the external environment. Figure 2 illustrates the muli-agent system architecture, created using the Prometheus Design Tool (PDT) [19], which is a graphical editor for specifying, designing and implementing intelligent agents.

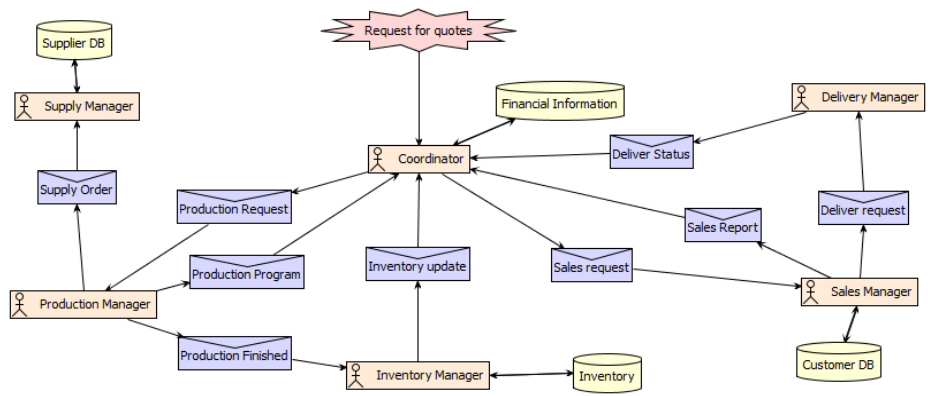

**Fig. 2.** Multi-agent system architecture.

Having separated agents allow us to reuse the system in parts, experiment with them in a wide range of real SCM applications, because every agent can be easily included or removed from the system without affecting the performance of the entire system. Although every agent has its own goals, the agents work in cooperation to reach the common global goal, which is to maximize the total profit. This goal can be divided in subgoals, such as maximizing the profit in sales, minimizing the prices when buying materials, minimizing the costs of storage, minimizing the penalizations for late deliveries, etc.

The following code fragment, written in Jason [20], shows the project where the multi-agent system is being implemented. It includes the six collaborative agents previously mentioned.

```
/* Jason Project */
MAS cMS {
    infrastructure: Centralised
    agents:
        coordinator;
```

```
        supply_Manager;
        production_Manager;
        inventory_Manager;
        delivery_Manager;
        sales_Manager;
}
```

This initial implementation was written in Jason, due to the possibility of using the internal actions implemented in Java to allow agents more autonomy to carry out their tasks, abilities for information management, and data processing. This also helps separating the abstraction level from the coordination, communication and deliberation. Additionally, this allows testing the system in a different implementation from the one that will be used for the TAC SCM experimentation platform, which uses Agentware.

The following Jason code fragment implements part of a plan for the *Coordinator Agent*, in which it receives a request. The code shows the declaration of the initial believes and rules; goals; and the implementation of the plans, which will be used to carry out the agent's tasks; it will also try to accomplish the goals specified at the beginning, and those that the other agents request.

```
// Agent coordinator in project CMS.mas2j
/* Initial beliefs and rules */
…
/* Initial goals */
…
/* Plans */
…
+!attendRequest(RQST) : true <-
    .evalProgram(ProdProgram);
    .send(delivery_Manager,achieve,evalDelivery(ProdSchedule));
    .updateProdEschedule(ProdProgram,ProdSchedule,RQST);
    .send(production_Manager,achieve,evalProdRequest(RQST,
        ProdProgram,TimeProduction));
    .send(production_Manager,achieve,prodRequest(RSQT,
        ProdProgram);
    .send(delivery_Manager,achieve,updateDelivery(ProdSchedule,
        TimeProduction).
…
```

The `attendeRequest(RQST)` plan takes a request, which is represented by the `RQST` parameter. The plan makes an evaluation of its production schedule, then it asks the `delivery_Manager` agent to evaluate the request according to its delivery schedule; and based on its answer it makes an update of that schedule. Finally, it asks the `production_Manager` agent to evaluate the request, and the `delivery_Manager` to estimate the delivery time; and with this information it makes the request that received initially.

## 3.1 Collaborative Agents

This subsection provides a brief description of the six collaborative agents that are part of the multi-agent system. They are implemented in Jason, following the Prometheus methodology.

**Coordinator Agent.** It is responsible for the communication with the external environment, such as the experimentation platform TAC SCM. It also coordinates the rest of the agents. Among its responsibilities are the following: update the inventory of materials and finished products; update the bank account status, receive offers from suppliers; receive requests and orders from clients; send offers to clients; send requests and orders to suppliers; share timetables of production and delivery of products with other agents; receive market and price reports; take a registry of requests, offers, orders, timetables of production and delivery of products, reports, and other information shared by the other agents; coordinate the performance of the entire system.

**Sales Manager Agent.** It is in charge of the product sales to the clients. It receives daily requests and orders from clients. For each of the client requests, this agent decides the price it will offer, through the prediction of prices. In the following section we consider some strategies that can be implemented for predicting prices of client orders. This agent must be in communication with the *Production Manager Agent*, in order to determine whether the production will be sufficient for satisfying the future client orders. The goal of this agent is to maximize the profit from the client orders.

**Supply Manager Agent.** This agent is in charge of generating requests for materials to the suppliers, considering their demand, the current level of use of materials, and the available stock. It should use strategies to send requests for materials, and make predictions in order to guarantee that there is enough stock, in order for the *Production Manager Agent* be able to manufacture products from the available materials. This agent has to carry out an analysis of the materials prices, and determine when to order them, in order to minimize the price for them. This analysis can be based on the prices paid recently, the current prices, and the prices provided by the market report. Once the *Supply Manager Agent* receives offers from the suppliers, it is in position to generate an order.

**Inventory Manager Agent.** It is responsible for receiving the materials from the suppliers; receiving finished products sent by the *Production Manager Agent*; sending materials for assembling products; delivering finished products to clients. This agent keeps track the stock of materials and products which are requested by the *Production Manager Agent* and the *Delivery Manager Agent*, respectively; and attempts to avoid that the inventory falls down a specified threshold, in order to satisfy the client demands. In order to minimize storage costs, this agent adjusts the limit of materials and products dynamically. It communicates with the *Production Manager Agent* and the

*Supply Manager Agent* through the *Coordinator*, in order to maintain the available stock of materials and products.

**Production Manager Agent.** It is responsible for programming the current production schedule and predicting the future production schedule. This agent can program its production schedule to satisfy its client because it has information about the client requests and orders from the *Sales Manager Agent*, and about the stock of materials in the inventory from the *Inventory Manager Agent*. In order to maximize its profit, it must be considered that each agent has a limited production capacity. This agent daily programs the production orders depending on the delivery dates, profit and availability of materials in stock, in order to proceed with more requests of materials if they are needed, through the *Supply Manager Agent*.

**Delivery Manager Agent.** The task of this agent is to deliver the products according to the client orders. In order to avoid penalizations for late deliveries, this agent schedules the active deliveries as soon as the products are ready in the production. This agent is in charge of reviewing the delivery orders, ordering them by delivery date, and delivers them according to the available products in the inventory.

## 4 Strategies for Agents

This section describes some strategies that the multi-agent system will implement. The strategies are focused on the demand side (sales strategies), and on the manufacture of products (production strategies).

### 4.1 Sales Strategies

The dynamic generation of prices has become an important concept, which regulates the relationships between sellers and their clients in e-commerce environments. The sales strategies with lists of fixed prices for all clients do not work anymore, because clients have the possibility of comparing prices in minutes by using different comparison Web sites. The sellers have to be capable of reacting immediately to the changes in the market situation, including changes in the demand volumes, the strategies of their competitors, and the client preferences; they also have to take into consideration their stock, their manufacturing capacity, as well as the relationships with their suppliers. These and other factors produce uncertainty in the commercial process. Online auctions have demonstrated to be the most efficient mechanisms of dynamic price generation, which allow sellers and buyers to agree on the prices when they participate in e-commerce activities.

There are different types of online auctions, which define different negotiation protocols between sellers and buyers; and one of the most relevant problems about them is the determination of the winner. Some examples of solutions that have been proposed to solve this problem can be found in [21] and [22]. In particular, the problem is to predict the prices that will be given to client orders, using the type of auction known as *first price sealed bid reverse auction*, in which the bid of every bidder is sent in a sealed envelope, and the lowest bid wins the auction. In the context of SCM, the following scenario takes place: a number of manufacturers offer their prices on a product that the client requested; without knowing among them the prices that each other offered. The client places the order to the manufacturer that offered the lowest price for the product. The ability of a manufacturer to predict the lowest price proposed by its opponents is crucial, because that way it can have a successful strategy that helps them maximize their profit.

The strategies that will be explored and implemented are described briefly in the following paragraphs. One strategy is to predict the probability that the winner price of the auction is on a specific interval, and place a bid according to the most probable price. Another strategy is to predict the prices based on the details of the client requests, the market situation, and the results of previous auctions, and place a bid according to the predicted price. The third strategy consists in predicting the highest and lowest prices of the client orders for each product, based on a time series for those prices, and place a bid between them. Another strategy is to model the competitor's behaviour and the prices that offers, and place a bid just below them.

In order to carry out the set of strategies proposed, it will be used statistical and learning techniques, such as neural networks and genetic programming. The learning methods can react to market irregularities more successfully, and provide more satisfactory results, in conditions of a dynamic SCM.

### 4.2 Production Strategies

One of the biggest problems of the supply chain is to determine when to order materials, in order to manufacture the products that have been requested by the clients. For example, if a manufacturer orders materials to receive them in the near future, this brings the benefit of storage costs reduction, because the materials are not in the storage place. On the other hand, this can bring uncertainty because it is unknown whether the materials will arrive on the specified date; the client demand is also unknown, and it is not possible to give an accurate prediction. The needs of production can be satisfied more adequately if the client demand is known in advance. Additionally to these problems, it is needed to consider the risks that exist in case the suppliers stop selling materials (due to their limited production capacity and the market competition), or delivering the materials after the date in which they were required.

In one of the strategies proposed, the agent can place a combination of orders with a long delivery and short delivery dates, in such a way that if the rest of the agents (that are competing) order materials with short delivery dates, then the supplier will have little free production capacity; that way the prices that correspond will be higher than those orders with long delivery date. In this strategy it must be observed the importance of an agent to be capable of automatically change its strategy dynamically; there will have to be an analysis of the strategies of the other competitor agents.

Another strategy should take into consideration the inventory level of raw materials as well as of finished products. For example, it can be maintained an inventory level based on the expected client demand, for a number of specific days, in such a way that the limit of the inventory can be calculated from the requested products in a specific number of days. Another strategy related with the inventory would be to maintain it in a specific level, and order materials only after having won client orders; that way it also reduces storage costs because as soon as materials arrive, they go to production; and as soon as they are manufactured, they are delivered to the clients. An adequate prediction of the suppliers' capacity and delays in the deliveries would help to solve the problem. There will have to be strategies that allow dynamically adjusting the production schedules, taking into consideration the factors mentioned.

## 5 Conclusions and Future Work

The Internet has changed the individual business processes, towards a more distributed business model, collaborative, in an e-commerce environment, and adaptable to competitive and changing market conditions. This paper presented a multi-agent system architecture for supply chain management, which explores different sales and production strategies and offers solutions in a distributed e-commerce environment. The system is designed to support different types of interfaces, which allow interoperating with other business models already developed.

There are six collaborative agents that compose the system (coordinator, sales manager, supply manager, inventory manager, production manager and delivery manager), which are implemented in Jason, following the Prometheus methodology. In order to show how the entire multi-agent system is being developed, an implementation of the coordinator agent was presented and explained, along with its architecture.

Future work includes the implementation and tests of the sales and production strategies in the experimentation platform TAC-SCM, in which the system can compete against other international agent systems, in order to measure its entire performance. Further work is also needed to individually test the implementation of the six collaborative agents with the strategies proposed.